\documentclass[jap,
                    a4,twocolumn
                    ]{revtex4}

\usepackage{epsfig}
\usepackage{amsmath}
\usepackage{setspace}
\usepackage{subfigure}
\usepackage{color}

\def\e{\begin{equation}}
\def\f{\end{equation}}

\begin{document}


\title{Broadband cloaking with volumetric structures composed of two-dimensional transmission-line networks}

\author{Pekka~Alitalo$^{1,2}$, Olli Luukkonen$^{1}$, Juan Mosig$^{2}$, Sergei~Tretyakov$^{1}$}

\affiliation{$^1$~Department of Radio Science and Engineering /
SMARAD Center of Excellence\\ TKK Helsinki University of
Technology \\P.O. Box 3000, FI-02015 TKK, Finland\\$^2$~Laboratory of Electromagnetics and Acoustics (LEMA), Ecole Polytechnique Federale de Lausanne (EPFL)\\
Batiment ELB, Station 11, CH-1015 Lausanne, Switzerland\\
}


\maketitle

\parskip 7pt

\vspace{0.5cm}
\begin{center}
\section*{Abstract}
\end{center}

The cloaking performance of two microwave cloaks, both based on
the recently proposed transmission-line approach, are studied
using commercial full-wave simulation software. The cloaks are
shown to be able to reduce the total scattering cross sections of
metallic objects of some restricted shapes and sizes. One of the
studied cloaks is electrically small in diameter, and the other is
electrically large, with the diameter equal to several
wavelengths.

\noindent \textit{Keywords:} Transmission-line network; scattering
cross section; electromagnetic cloak.

\medskip
\noindent \textit{PACS} 41.20.Jb, 84.40.Xb

\vspace{1.6cm}

\section{Introduction}

Cloaking, i.e., the reduction of an object's total scattering
cross section for arbitrary incidence directions, has recently
gained wide interest in the literature, see the review paper by
Al$\rm{\grave{u}}$ et al.~\cite{Alu_review} for a representative
list of references and explanations of the main cloaking
principles. There exists several different approaches to cloaking,
such as the coordinate-transformation
approach~\cite{Leonhardt,Pendry,Schurig} and the so-called
plasmonic cloaking~\cite{Alu_review,Alu}.

In addition to the aforementioned cloaking techniques, an
alternative aproach has been proposed
recently~\cite{Alitalo_cloak_TAP}. This approach is based on the
use of transmission-line networks that are coupled with the
surrounding medium, and it has been studied in our recent
papers~\cite{Alitalo_cloak_TAP,Alitalo_cloak_iWAT08,Alitalo_cloak_URSI08,Alitalo_cloak_META}.
In this paper we study two cloaks that are based on the designs
presented in~\cite{Alitalo_cloak_iWAT08,Alitalo_cloak_URSI08} and
analyse their cloaking performance in free space. Both studied
cloaks are cylindrically shaped and consist of periodically
stacked two-dimensional networks of transmission lines. One of the
cloaks is electrically small, with the diameter of the cloaked
region about one fourth of the wavelength. The other cloak is
electrically large, with the diameter of the cloaked region being
approximately four wavelengths. Both cloaks are studied under free
space conditions with TE-polarized plane wave illumination.

To get more insight on the applicability of these types of cloaks
in realistic antenna applications (ideal plane-wave illumination
is not a good approximation for many practical situations), we
apply the designed electrically small cloak in a situation where a
metallic object is placed near a radiating dipole antenna, thus
affecting the radiation pattern of the antenna.

\section{Electrically large cloak}

\begin{figure} [h!]
\centering {\epsfig{file=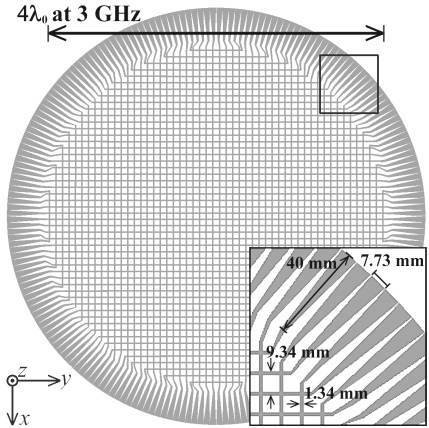,
width=0.35\textwidth}} \caption{Electrically large cloak. The
network of parallel-strip transmission lines lies in the
$xy$-plane. The inset shows a magnification of a part of the cloak
with dimensions.} \label{bigcloak}
\end{figure}

The electrically large cloak is similar to the one recently
presented in~\cite{Alitalo_cloak_URSI08}. We use a two-dimensional
transmission-line network with the shape as illustrated in
Fig.~\ref{bigcloak}. The transmission lines are composed of
parallel metal strips and the coupling between the network and the
free space surrounding the network is achieved by using a
``transition layer.'' This layer is formed by gradually enlarging
the transmission line strips to cover the whole interface between
the network and free space~\cite{Alitalo_cloak_TAP}. In this case
the two-dimensional network lies in the $xy$-plane, and the cloak
structure can be made volumetric by stacking these networks on top
of each other along the
$z$-direction~\cite{Alitalo_cloak_TAP,Alitalo_cloak_iWAT08,Alitalo_cloak_URSI08,Alitalo_cloak_META}.

The objects that can be cloaked with this set of networks include
two-dimensional arrays or three-dimensional meshes of metal or any
other material. Any type of object that can fit in the space that
is inside the cloak but outside the sections of transmission
lines, can be cloaked with this approach. The cloaking effect is
naturally best for cloaking of metallic objects since alone these
objects scatter significantly.

The cloaking efficiency of the electrically large cloak of
Fig.~\ref{bigcloak} is studied using a commercial full-wave
simulation software Ansoft HFSS~\cite{Ansoft}. To simplify the
simulations, we consider the cloak structure to be infinitely
high, i.e., there is an infinite number of the two-dimensional
cloaks stacked on top of each other creating a volumetric
structure. The period $d$ of the two-dimensional transmission-line
networks, from which the volumetric cloak is constructed from, is
8~mm. We can hence assume that the networks have isotropic
propagation properties at least up to the frequency of
3~GHz~\cite{Alitalo_cloak_TAP}.

To improve the cloaking efficiency of the previously studied
structure~\cite{Alitalo_cloak_URSI08} at the design frequency of
3~GHz, the width of the parallel strip transmission lines is tuned
from 1.38~mm~\cite{Alitalo_cloak_URSI08} to 1.34~mm with the
separation (i.e., height) of the transmission lines being 2~mm in
the $z$-direction. This tuning can be considered as tuning of the
characteristic impedance of the network itself, as explained
in~\cite{Alitalo_cloak_TAP}. With this tuning the
transmission-line networks are matched better to the free-space
impedance and thus the cloaking effect is improved. The
transmission lines of the transition layer have equal width and
height and the length of these lines is equal to 40~mm, see
Fig.~\ref{bigcloak}. At the interface between the network and free
space, the width and height of the transmission lines of this
transition layer are equal to 7.73~mm. This value is determined
simply by the circumference of the cloak structure at this
interface and by the number of transmission lines in the
transition layer (circumference divided by the number of
transmission lines).

The transmission-line strips are modelled in the simulation
software as infinitely thin sheets of perfectly conducting
material (PEC). As a reference object we use the object that we
want to cloak, which in this case is a two-dimensional array
consisting of infinitely long PEC rods with a square cross section
of 4~mm~$\times$~4~mm. The array lies in the $xy$-plane, i.e., the
rods are directed along the $z$-axis. This array has a cylindrical
shape so that it fills the available space inside the cloak
network almost entirely and it has the same period as the network,
i.e., $d=8$~mm. There are 51 PEC rods along the diameter of this
cylindrical array, which means that the cloaked object's
electrical size is approximately $4\lambda_0$ at 3~GHz.

The cloak operation is studied by simulating the scattering cross
sections (SCS) of the reference object with and without the cloak
to all directions in the $xy$-plane. The SCS is simulated by
illuminating the simulation models with TE-polarized plane waves
(electric field $E$ parallel to the $z$-axis) and then calculating
the scattered power in the far field. By integrating the SCS over
the angle $\phi$ which lies in the $xy$-plane, we obtain the total
SCS of the two cases. To give a simple representation of the
results, we normalize the total SCS of the cloaked object to the
total SCS of the uncloaked object. Thus, the resulting normalized
total SCS ($SCS_{\rm tot,n}$) is smaller than 1 when reduction of
the total scattering cross section is achieved.

\begin{figure} [b!]
\centering {\epsfig{file=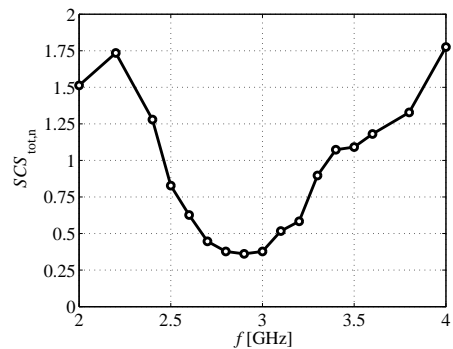,
width=0.4\textwidth}} \caption{Total scattering cross section of
the cloaked object, normalized to the total scattering cross
section of the uncloaked object. The illuminating plane wave
travels in the $+x$-direction.} \label{bigcloak_SCS}
\end{figure}

In Fig.~\ref{bigcloak_SCS} $SCS_{\rm tot,n}$ versus the frequency
is shown for the electrically large cloak. $SCS_{\rm tot,n}$ is
smaller than 1 in a relative bandwidth of approximately 15
percent, with the center frequency at 2.9~GHz. The minimum value
of $SCS_{\rm tot,n}$ is obtained at the frequency of 2.9~GHz,
where $SCS_{\rm tot,n}\approx 0.36$, i.e., the total scattering
cross section of the reference object is reduced by approximately
64 percent. At frequencies below approximately 2.5~GHz and above
3.3~GHz, the phase velocity mismatch between the cloak and the
free space causes an increase in the cloak's scattering even above
the level of the bare reference object, as
expected~\cite{Alitalo_cloak_TAP}.

\section{Electrically small cloak}

\begin{figure} [b!]
\centering \subfigure[]{\epsfig{file=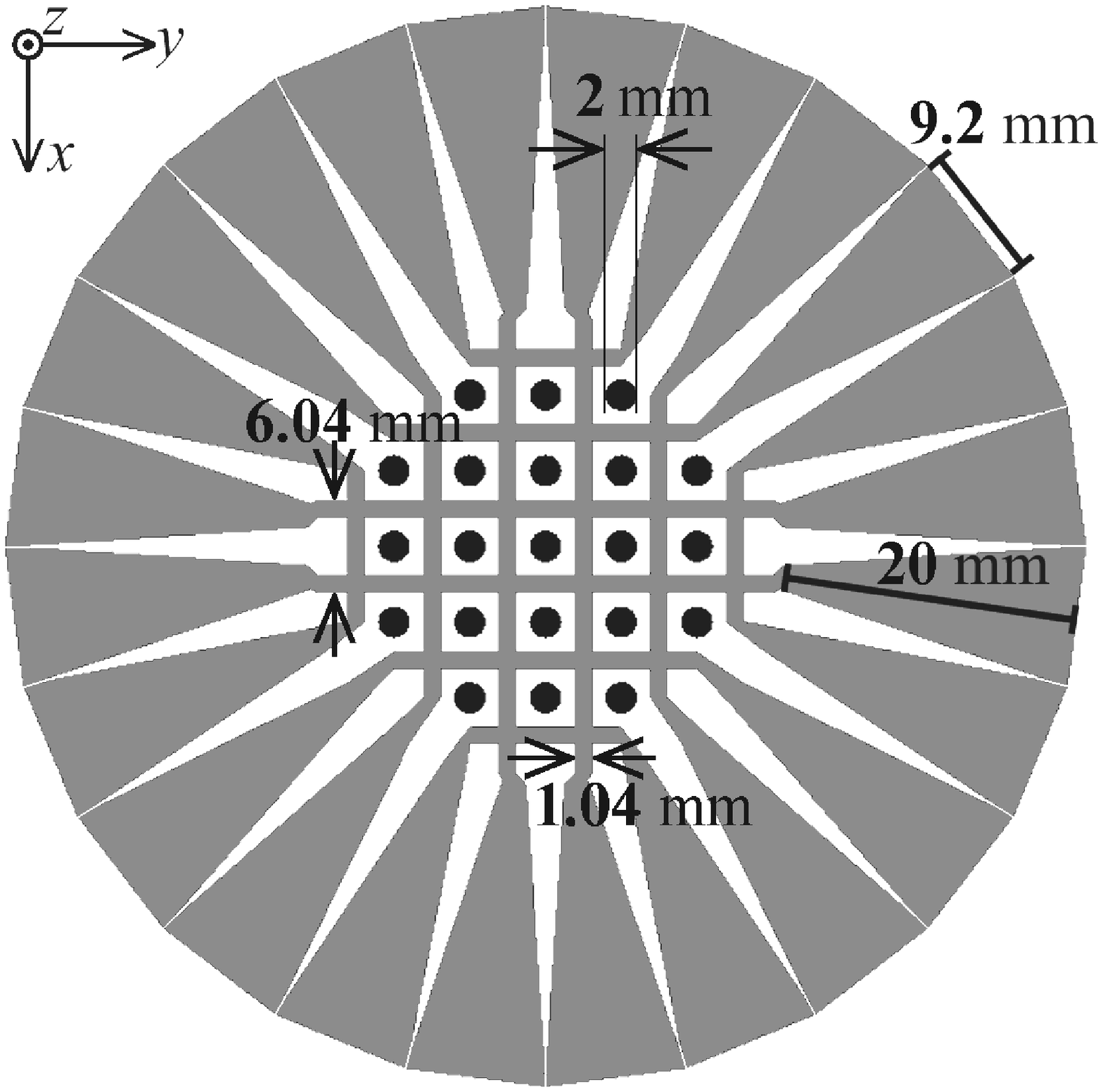,
width=0.3\textwidth}} \subfigure[]
{\epsfig{file=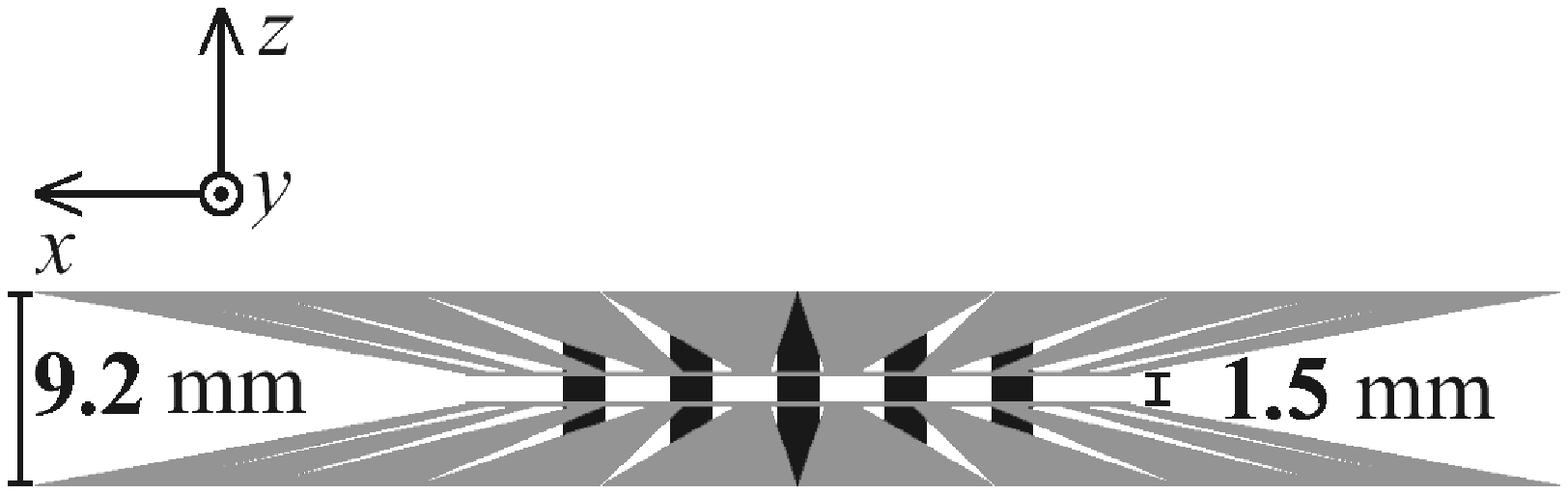, width=0.3\textwidth}}
\caption{Electrically small cloak. Grey illustrates the
transmission lines of the cloak and black the reference object.}
\label{smallcloak}
\end{figure}

The electrically small cloak studied here is a modification of a
similar cloak presented in~\cite{Alitalo_cloak_iWAT08}. Here we
tune the operation frequency of the cloak from
2~GHz~\cite{Alitalo_cloak_iWAT08} to 3~GHz, which was the
operation frequency also in the previous example. This can be done
simply by keeping the cloak structure the same as
in~\cite{Alitalo_cloak_iWAT08} but just decreasing the period of
the transmission-line networks to obtain the same electrical size
(diameter) of the cloak at a higher frequency. The dimensions of
the transmission lines are also tuned accordingly to obtain
optimal impedance matching at a higher frequency. This tuning is
done with full-wave simulation software~\cite{Ansoft}, by changing
the width of the strips gradually and analyzing the total
scattering cross section. The resulting cloak structure, with the
period $d$ of the network reduced from 8~mm to 5~mm and the length
of the transmission lines in the transition layer reduced from
40~mm to 20~mm, is shown in Fig.~\ref{smallcloak}. The resulting
network has a diameter of $6d=30$~mm, while the diameter of the
cloaked object is $D=22$~mm, i.e., $0.22\lambda_0$ at 3~GHz.

Again, as in the case of the electrically large cloak, the
dimensions of the transition layer depend on the circumference of
the transition layer and the number of transmission lines in this
layer, now resulting in the width and height of 9.2~mm at the
interface with free space. The transmission lines composing the
network itself are again realized as parallel metal strips
(modelled by infinitely thin PEC strips in the simulation
software), with the width and height equal to 1.04~mm and 1.5~mm,
respectively. The reference object in this case is a
two-dimensional array of infinitely long PEC rods, now having a
circular cross section with the diameter of 2~mm, see
Fig.~\ref{smallcloak}.

\begin{figure} [b!]
\centering {\epsfig{file=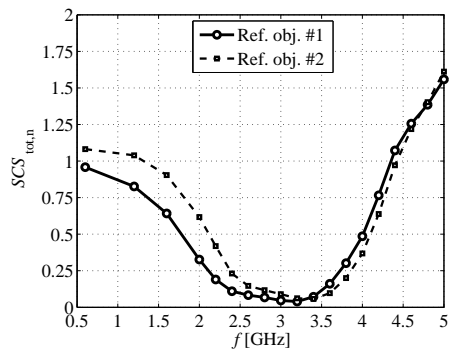,
width=0.4\textwidth}} \caption{Total scattering cross section of
the cloaked object, normalized by the total scattering cross
section of the uncloaked object. Ref. obj.~\#1: two-dimensional
array of PEC rods. Ref. obj.~\#2: two-dimensional array of PEC
rods connected with periodically inserted discs of PEC. The
illuminating plane wave travels in the $+x$-direction.}
\label{smallcloak_SCS}
\end{figure}

The scattering simulations are conducted as described in the
previous section, i.e., a TE-polarized plane wave travelling in
the $+x$-direction illuminates the structures which are modelled
as infinitely periodic in the $z$-direction. The resulting
$SCS_{\rm tot,n}$ versus the frequency is plotted in
Fig.~\ref{smallcloak_SCS} (solid line), showing the relative
cloaking bandwidth of approximately 158 percent, with the center
frequency being at 2.425~GHz. The best cloaking performance with
$SCS_{\rm tot,n}\approx 0.04$ is obtained at the frequency of
3.2~GHz. Therefore, at this frequency the total scattering cross
section of the reference object is reduced by 96 percent.

The isotropy of the cloak in the $xy$-plane is confirmed by
conducting the same scattering simulations as above for different
incidence angles in the $xy$-plane. Due to the symmetry of the
cloak, it is enough to study incidence angles between $0^0$ and
$45^0$. We carried out the scattering simulations between these
incidence angles, with the step of $5^0$, at the frequency of
3.2~GHz. The resulting $SCS_{\rm tot,n}$ varied approximately
between the values of 0.03 and 0.045.

\section{Cloaking an object near a dipole antenna}

\begin{figure} [b!]
\centering \subfigure[]{\epsfig{file=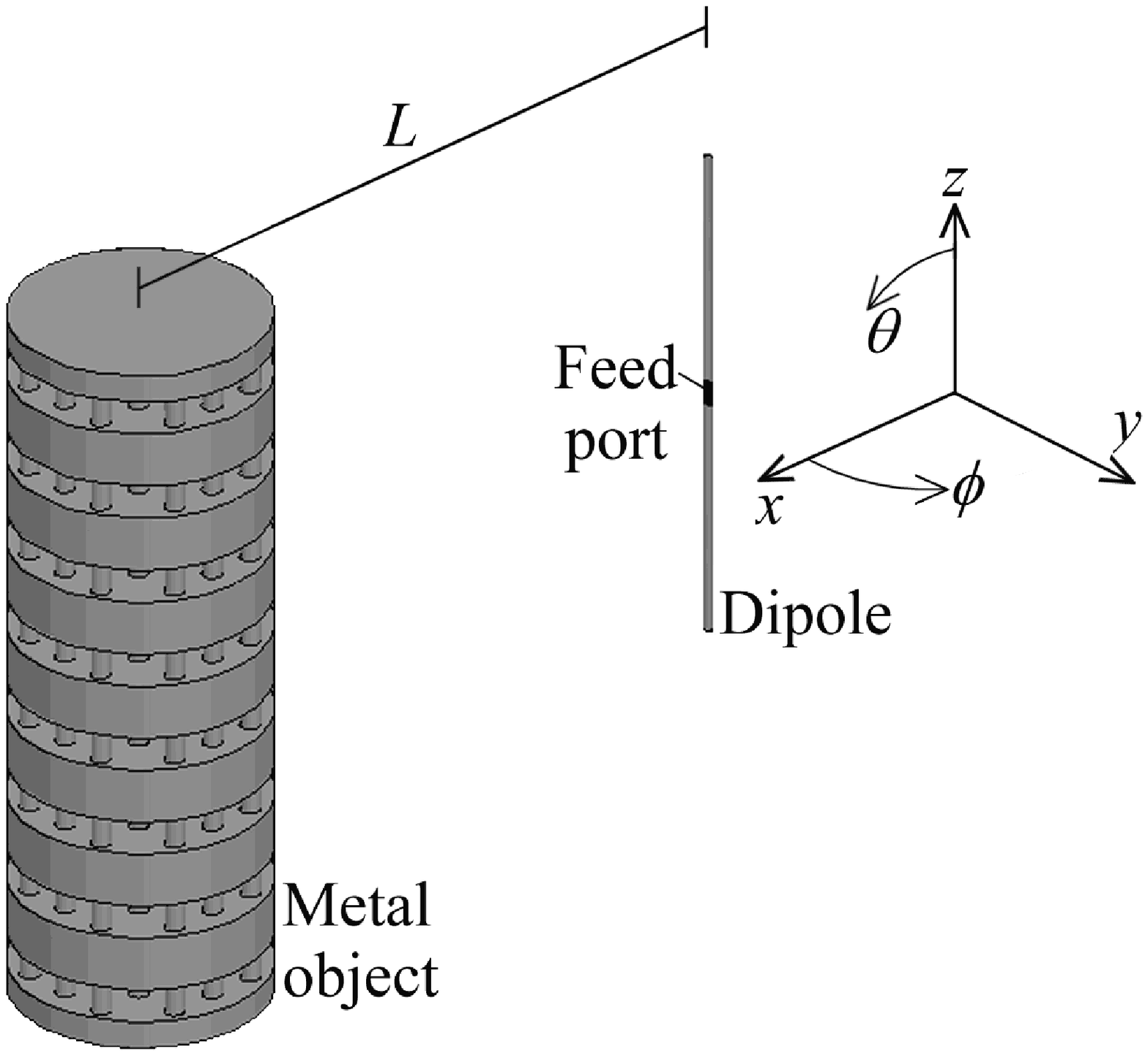,
width=0.32\textwidth}} \subfigure[] {\epsfig{file=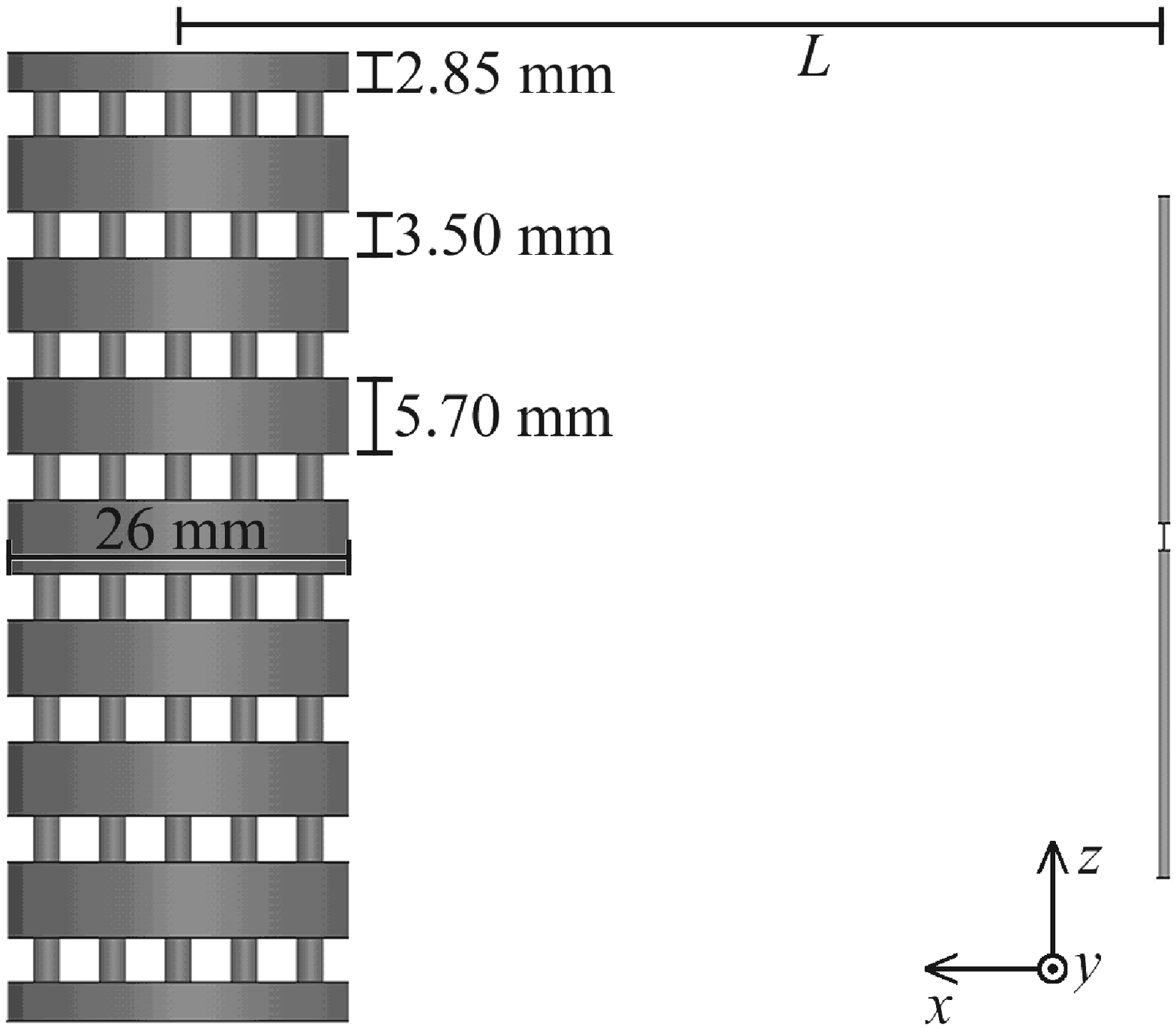,
width=0.28\textwidth}} \caption{Illustration of the simulation
model with a dipole antenna and a metal (PEC in this case) object
blocking the antenna radiation. The dipole length is 50~mm
($\lambda_0/2$ at 3~GHz).} \label{smallcloak_dipole}
\end{figure}

Plane-wave illumination is not a good approximation in many
realistic situations where cloaking would be beneficial.
Especially in antenna applications this will most probably be an
important issue. To demonstrate the robustness of our cloaking
approach with respect to this problem, we simulate a finite-sized
cloak near a dipole antenna. The cloak is used to ``hide'' a
metallic object that blocks the antenna's radiation in certain
directions. See Fig.~\ref{smallcloak_dipole} for the illustration
of the problem. The metallic object is such that we can use the
electrically small cloak, studied in the previous section, to
cloak it. Instead of a two-dimensional array of PEC rods we choose
to cloak a more realistic structure, that could be used as a
support strut, etc. In order to make the PEC rods of the reference
object connected to each other, we introduce PEC discs of a finite
height periodically connected to the array of PEC rods (same as in
Fig.~\ref{smallcloak}), as illustrated in
Fig.~\ref{smallcloak_dipole}.

First, to study the effect of changing the reference object, we
repeat the scattering simulations done in the previous section for
this new reference object (modelled as an infinitely high
structure, due to the periodicity of the simulation model
discussed in the previous section). The results are shown in
Fig.~\ref{smallcloak_SCS} (dashed line), demonstrating that the
cloaking effect remains the same, just a slight shift in the
frequency of the optimal cloaking performance is observable.

Two different cases of the problem presented in
Fig.~\ref{smallcloak_dipole} are studied: the distance $L$ between
the dipole and the metallic object being $L=100$~mm and $L=50$~mm.
This is to study the operation of the cloak in the near-field
region of the antenna (note that the wavelength at 3~GHz is equal
to 100~mm). Also, moving the object closer to or further from the
antenna changes the incidence angle of the fields that impinge on
the cloak in both $xy$- and $xz$-planes. One purpose of these
simulations is to study how sensitive the cloak is to changes in
the incidence angle $\theta$.

To cloak this metal object, we have used the electrically small
cloak studied in the previous section, but in this case the cloak
has a finite height with eight periods of the cloak in the
$z$-direction. As can be seen from Fig.~\ref{smallcloak_dipole},
the metal object requires placement of eight two-dimensional
transmission-line networks one upon another in order to cover the
whole object in the $z$-direction. The height of the metal object
is $8\times 9.2$~mm$=73.6$~mm. The transmission-line networks of
the cloak structure run in the space between the PEC discs, where
there are only the vertical PEC rods, similar to
Fig.~\ref{smallcloak}.

\begin{figure} [t!]
\centering \subfigure[]{\epsfig{file=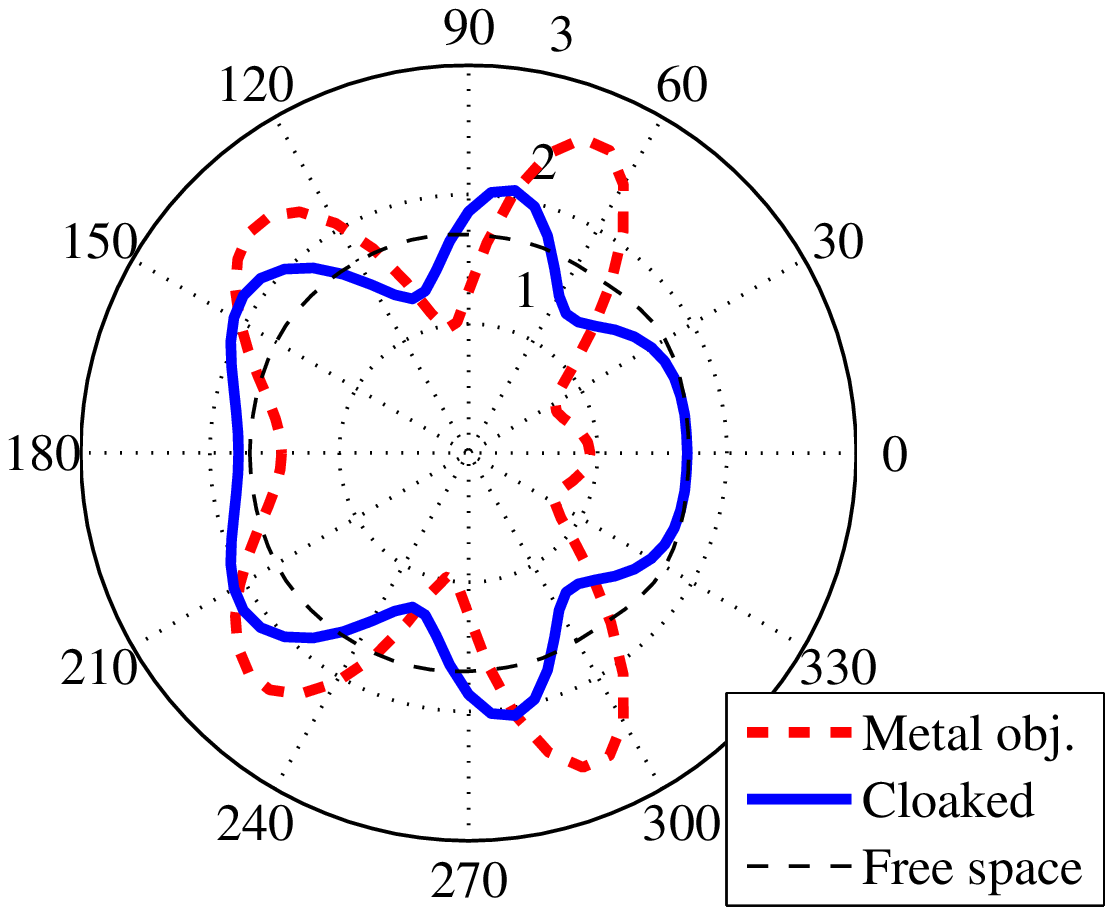,
width=0.41\textwidth}} \subfigure[]
{\epsfig{file=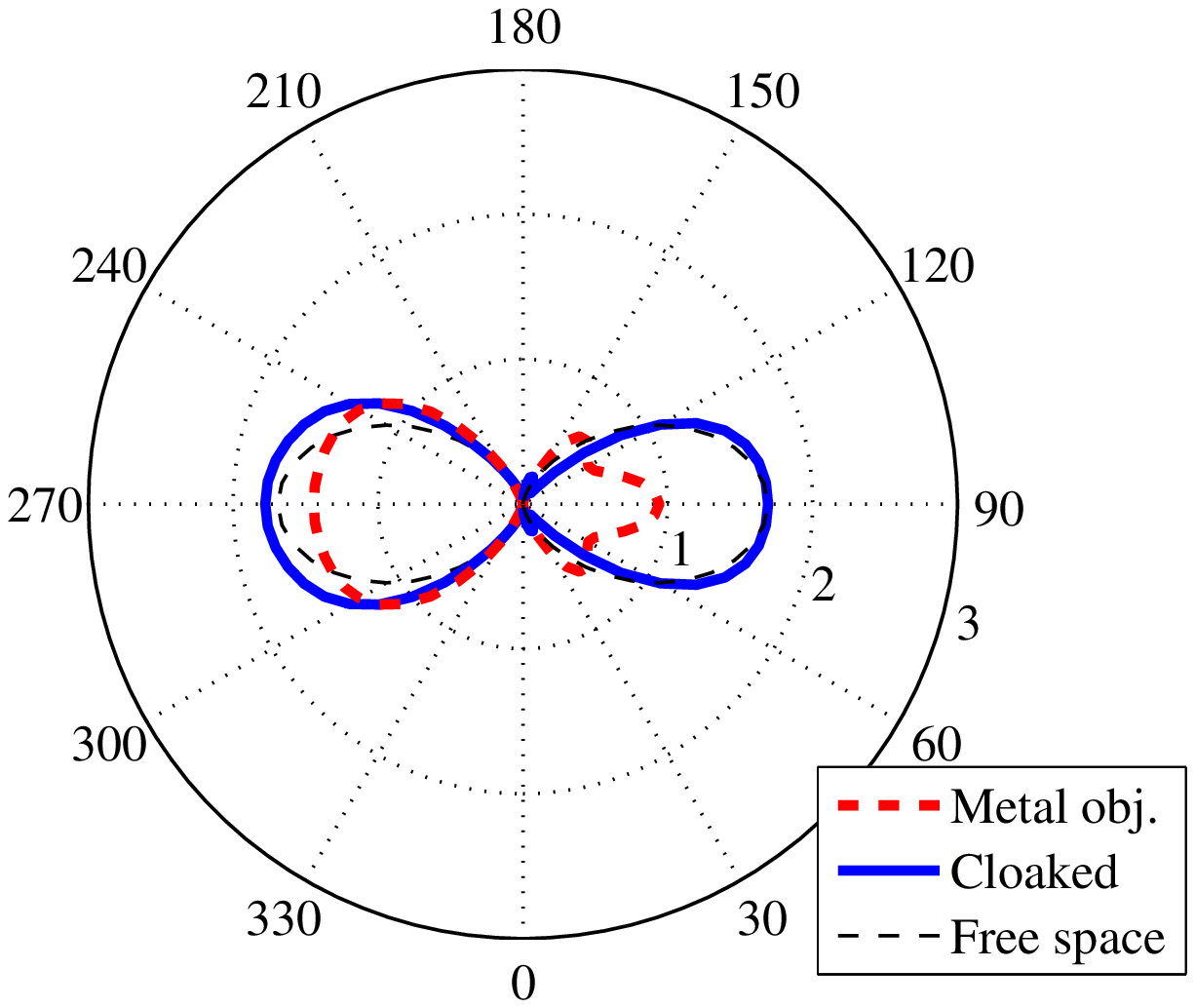, width=0.37\textwidth}}
\caption{Simulated directivity patterns (linear scale) for
$L=100$~mm at the frequency $f=3.1$~GHz. (a) H - plane
($\theta=90^0$). (b) E - plane ($\phi=0^0$).}
\label{smallcloak_v1_D31}
\end{figure}

Three different scenarios were simulated for both values of $L$:
1)~dipole alone in free space, 2)~dipole with the metallic object
at distance $L$ away from the dipole, and 3) dipole with the
cloaked metal object. The easiest way to analyse the results is to
study the radiation patterns in the $E$- and $H$-planes and
compare the different scenarios with each other. We have run
simulations studying several frequencies, and for both values of
$L$ the best directivity patterns are obtained at 3.1~GHz. The
goal here is to have as good agreement with the free space
scenario and the cloaked scenario as possible, in the forward and
backward directions ($\phi=0^0$, $\theta=90^0$ and $\phi=180^0$,
$\theta=90^0$). See Fig.~\ref{smallcloak_v1_D31} for the results
for the case of $L=100$~mm, which demonstrates that the metal
object alone is causing massive deformation of the radiation
pattern. When the object is ``cloaked,'' the directivity pattern
resembles the free space situation much better, both in $E$- and
$H$-planes.

\begin{figure} [t!]
\centering \subfigure[]{\epsfig{file=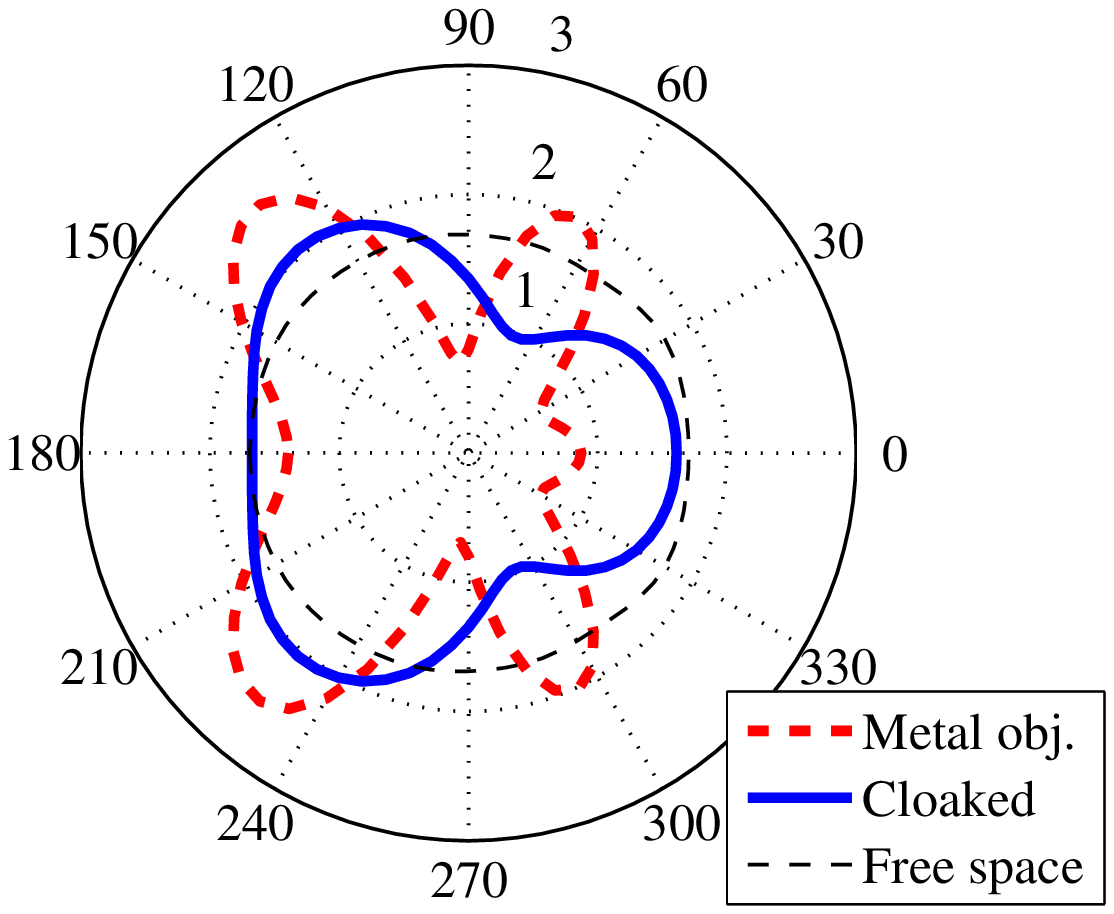,
width=0.41\textwidth}} \subfigure[]
{\epsfig{file=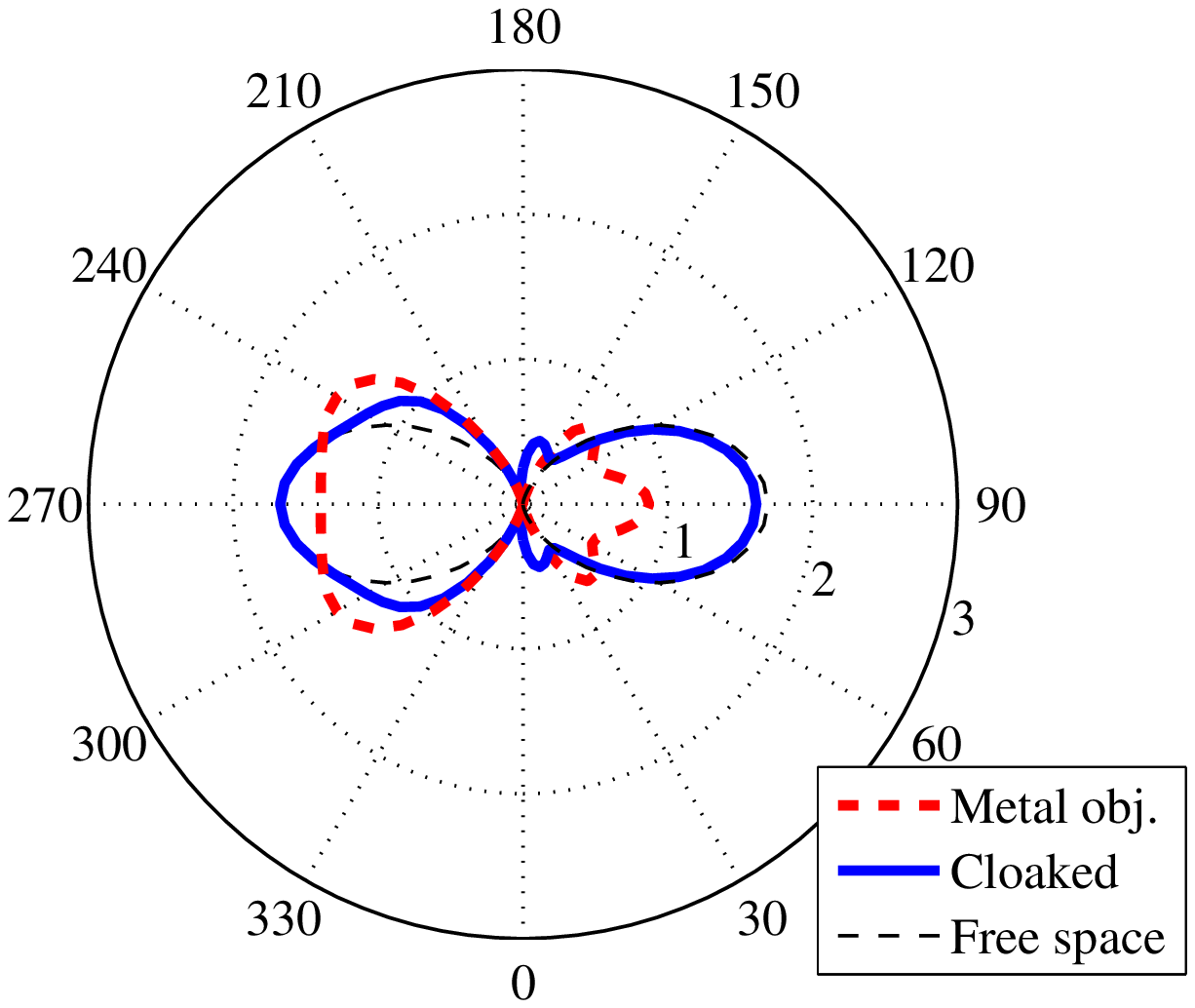, width=0.37\textwidth}}
\caption{Simulated directivity patterns (linear scale) for
$L=50$~mm at the frequency $f=3.1$~GHz. (a) H - plane
($\theta=90^0$). (b) E - plane ($\phi=0^0$).}
\label{smallcloak_v2_D31}
\end{figure}

For the case of $L=50$~mm, see Fig.~\ref{smallcloak_v2_D31}. In
this case the cloak cannot preserve the ideal directivity pattern
as well as in the previous case. This is expected since the cloak
is now in the near-field region of the dipole. Still, for both
$H$-plane and $E$-plane patterns the radiation in the direction of
the object ($\phi=0^0$, $\theta=90^0$) is almost ideal, whereas
with the metal object alone, the radiated power in this direction
is strongly mitigated.

\section{Conclusions}

We have studied two electromagnetic cloaks, both composed of
layered two-dimensional transmission-line networks. The cloaking
performance is studied using a commercial full-wave simulation
software. The performance of both cloaks is first evaluated for
ideal plane-wave illumination. The other cloak, which is
electrically smaller and exhibits larger bandwidth, is simulated
also in an antenna scenario, where a metal object blocking the
antenna radiation is cloaked.

\section*{Acknowledgements}

This work has been partially funded by the Academy of Finland and
TEKES through the Center-of-Excellence program and partially by
the European Space Agency (ESA--ESTEC) contract no. 21261/07/NL/CB
(Ariadna program). P.~Alitalo acknowledges financial support by
the Finnish Graduate School in Electronics, Telecommunications,
and Automation (GETA), the Emil Aaltonen Foundation, and the Nokia
Foundation.



\begin{thebibliography}{99}


\bibitem{Alu_review}
A. Al$\rm{\grave{u}}$ and N. Engheta, ``Plasmonic and metamaterial
cloaking: physical mechanisms and potentials,'' \textit{Journal of
Optics A: Pure and Applied Optics}, vol.~10, 093002, 2008.

\bibitem{Leonhardt}
U. Leonhardt, ``Optical conformal mapping,'' \textit{Science},
vol.~312, pp.~1777--1780, 2006.

\bibitem{Pendry}
J.B. Pendry, D. Schurig, and D.R. Smith, ``Controlling
electromagnetic fields,'' \textit{Science}, vol.~312,
pp.~1780--1782, 2006.

\bibitem{Schurig}
D. Schurig, J.J. Mock, B. J. Justice, S.A. Cummer, J.B. Pendry,
A.F. Starr, and D.R. Smith, ``Metamaterial electromagnetic cloak
at microwave frequencies,'' \textit{Science}, vol.~314,
pp.~977--980, 2006.

\bibitem{Alu}
A. Al$\rm{\grave{u}}$ and N. Engheta, ``Achieving transparency
with plasmonic and metamaterial coatings,'' \textit{Phys.~Rev.~E},
vol.~72, 016623, 2005.





%
%
%
%





\bibitem{Alitalo_cloak_TAP}
P.~Alitalo, O.~Luukkonen, L.~Jylh\"a, J.~Venermo, and
S.~A.~Tretyakov, ``Transmission-line networks cloaking objects
from electromagnetic fields,'' \textit{IEEE Trans. Antennas
Propagat.} vol.~56, pp.~416--424, 2008.

\bibitem{Alitalo_cloak_iWAT08}
P. Alitalo and S. Tretyakov, ``Cylindrical transmission-line cloak
for microwave frequencies,'' \textit{Proc. 2008 IEEE International
Workshop on Antenna Technology}, 4-6 March 2008, Chiba, Japan,
pp.~147--150.

\bibitem{Alitalo_cloak_URSI08}
P. Alitalo and S. Tretyakov, ``On electromagnetic cloaking --
general principles, problems and recent advances using the
transmission-line approach,'' \textit{Proc. 2008 URSI General
Assembly}, 7-16 August 2008, Chicago, USA, p.~B01p9 (invited).

\bibitem{Alitalo_cloak_META}
P. Alitalo, S. Ranvier, J. Vehmas, and S. Tretyakov, ``A microwave
transmission-line network guiding electromagnetic fields through a
dense array of metallic objects,'' \textit{Metamaterials}, in
press. Preprint http://arxiv.org/abs/0805.4055.


\bibitem{Ansoft}
The homepage of Ansoft corporation, http://www.ansoft.com/.




\end{thebibliography}
\end{document}